%% file: skeleton.tex
\documentclass{PoS}

\input{tikz}

\title{$V$ and heavy flavour production: status and recent theoretical developments}

\ShortTitle{Heavy flavour associated production}

\author{\speaker{Davide Napoletano}\\
  IPhT, CEA Saclay, CNRS, Universit\'e Paris-Saclay,
  F-91191 Gif-sur-Yvette cedex, France.\\
  E-mail: \email{davide.napoletano@ipht.fr}}

\abstract{In this talk, we review the status of the associated
  production of a vector boson and heavy flavour quarks, and some
  recent theoretical developments. While some aspects of these
  multiscale processes are not yet completely under control, a general
  understanding on how to treat them is emerging, thanks to many
  efforts to match a completely massless, with a completely massive approach.}

\FullConference{7th Annual Conference on Large Hadron Collider Physics - LHCP2019\\
  20-25 May, 2019\\
  Puebla, Mexico}

\begin{document}

The associated production of a vector boson ($V=W/Z/\gamma*$) or
a Higgs boson with heavy flavour quarks presents many interesting
facets. From the experimental stand point, it constitutes irreducible
backgrounds for many Standard Model measurements and beyond the
Standard Model searches alike. From the theoretical stand point, it  
represents an important case study for multiscale problems.
Indeed, as heavy quarks are mostly produced from a gluon splitting, a
typical situation one may be in is
\begin{center}
  \begin{tikzpicture}[line width=0.9 pt, scale=1.1]
    \fill[pattern=north west hatch, hatch distance=7pt, hatch thickness=1pt, pattern color=black!30!white, opacity=.6,draw] 
    (0,0) circle [radius=0.7];
    \node[opacity=1] (0,0) {${\cal M}$};
    \draw[psgluon,opacity=0.6] (-2,0.3) -- (-1.3,0.3) ;
    \draw[lparticle,opacity=0.6] (-1.3,0.3) -- (-.7,0) ;
    \draw[lparticle,opacity=0.6] (-0.7,.6) -- (-1.3,0.3) ;
    \node[text width=9cm] at (2.8,0) {
      \begin{minipage}{\textwidth}
        \begin{equation}
          \label{eq:1}
          \sim \quad \!\!\int^{m_V^2}\frac{{\text{d} t}}{t-m_b^2}
          \sim \log\frac{m_V^2}{m_b^2}
        \end{equation}
      \end{minipage}
    };
  \end{tikzpicture}
\end{center}
where $m_V$ in this case is just a place holder for, some, mass-like,
hard scale.
As it can be seen from Eq.~(\ref{eq:1}), depending on the relative
size of $m_V$ and $m_b$ one can get contributions that are enhanced by
powers of the corresponding logarithm.

At this point, depending on the case at hand, one has the possibility
of making two choices. One possibility consists in accepting the
appearing logarithms (assuming they are small)
as a fact of life to retain heavy quark mass
effects. In order for this approach to work, one needs to define a
scheme in which, with respect to QCD evolution, the mass of the heavy
quark is effectively infinite, and thus decoupled. Such a scheme is
commonly referred to as the Four Flavour Scheme (4FS) in the case of
$b$-quarks, and clearly is
a valid approximation when the scale of interest does not exceed by
much the mass of the heavy quark.
Alternatively, one can decide to neglect all mass effects,
transforming the logarithm in Eq.~(\ref{eq:1}) in a collinear
divergence, and resum to all orders the contributions coming from it,
in the same way as for all other light quarks, effectively defining a
$b$-quark PDF. Such a scheme is called a Five Flavour scheme (5FS) and
it is expected to be valid in the region where $m_V\gg m_b$.

Comparisons between these two approaches for the production of
$b$-quarks in association with a vector boson, have shown that, overall,
the most accurate 5F prediction seems to favour data better. This has
been supported by various inclusive calculations aiming at matching
these two schemes: at scales relevant at the LHC, logarithmic effects
are generally more important than mass power
corrections\cite{Forte:2015hba,Forte:2016sja,Krauss:2016orf,Forte:2018ovl}.
Nevertheless, there are differential observables for which including
mass effects may still have an important impact. For these reasons
various recent attempts to consistently include both the resummation
of high energy logarithms and mass effects have been made.

\begin{figure}[t]
  \centering{
    \includegraphics[width=0.99\linewidth]{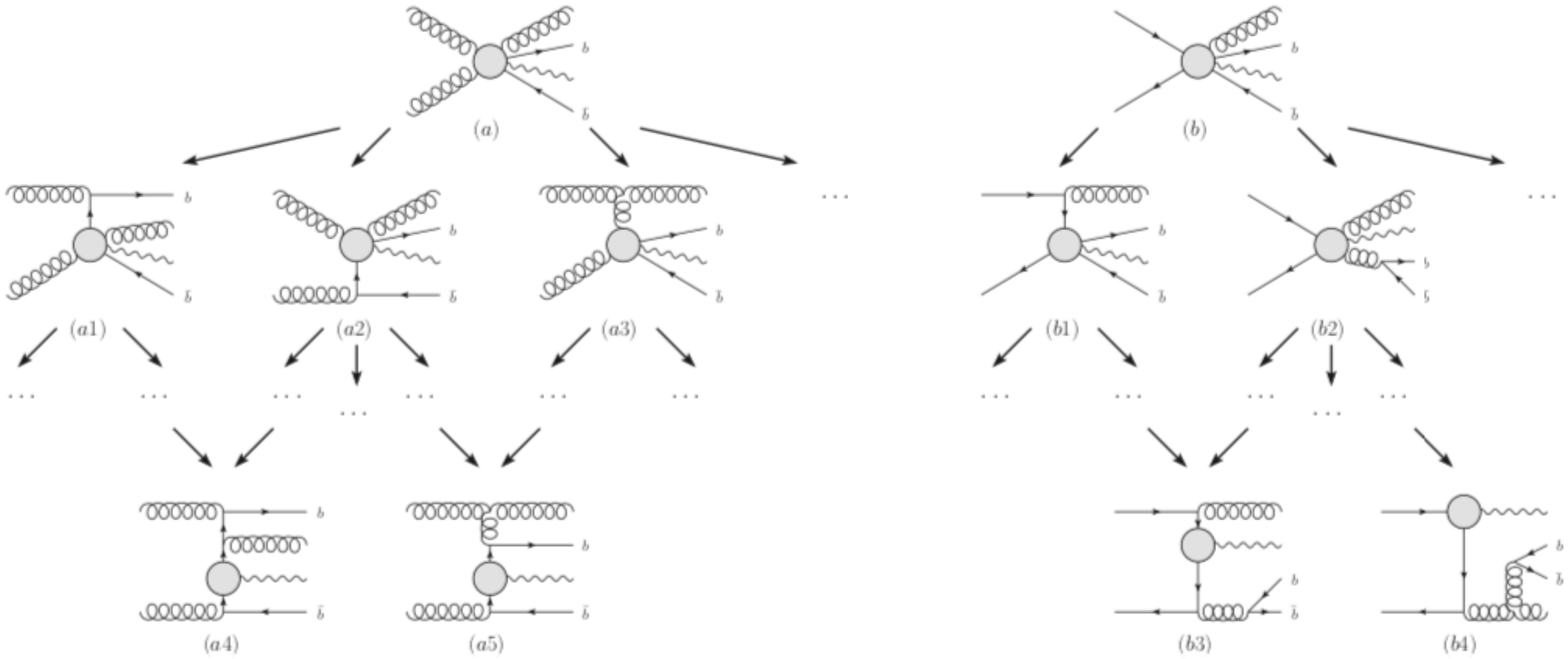}
  }
  \caption{\label{fig:mjm-scheme} Pictorial description of the
    clustering for one the various processes contributing to the
    production of a $Z$ boson in association with heavy flavours.}
\end{figure}

Although not discussed in this talk, among the various efforts it is
worth noting Ref.~\cite{Bagnaschi:2018dnh}. More recently a similar
approach in spirit have been developed in the context of the SHERPA
Monte Carlo generator, based on multi-jet
merging~\cite{Hoche:2019ncc}.
A diagrammatic representation of how the multi-jet merging algorithm
works in this case is presented in Fig.~\ref{fig:mjm-scheme}.

The working idea in this case is quite straightforward. Taking as an
example the production of a $Z$ boson plus a pair of $b$-quarks, one
starts by generating both an inclusive 5FS sample which corresponds to
$Z + \text{jets}$ with multi-jet merging, and a calculation in the
4FS, $Z+b\bar{b}$. The next step consists in processing the latter in the
same way as if it came from the former, by applying the normal
clustering procedure. This generates what the authors call the
\textit{direct} component. The last step (\textit{fusing}) consists in
removing from the 5FS sample all events that have a configuration
which can also be generated by the reclustered 4FS one. This creates
what the authors refer to as the \textit{fragmentation} component, and
one obtains the fused result by combining the two samples.

\begin{figure}[t]
  \centering{
    \includegraphics[width=0.99\linewidth]{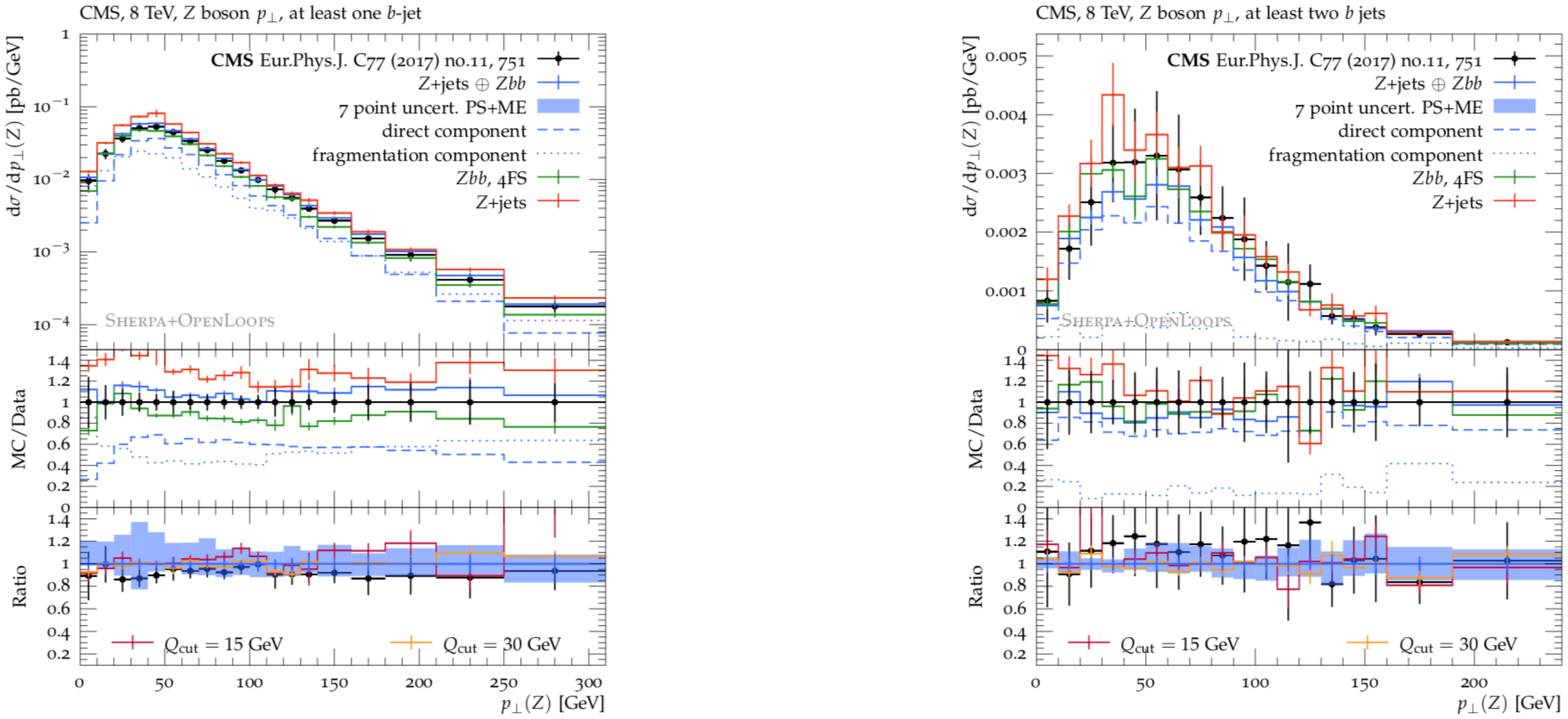}
  }
  \caption{\label{fig:mjm} Comparison of the multijet merging (fused)
    scheme with the standard 5FS as well as with the components of the
  merged results. The left panel refers to an event selection where
  one has at least one $b$ jet, while in the right panel, a selection
  with at least 2 $b$-jets is made.}
\end{figure}

The authors then go on and make some connection to the inclusive FONLL
method. Here it has to be noted that while indeed they are able to
construct some relevant terms necessary to construct the FONLL
matching, the accuracy they get is only NLO plus that of the parton shower
(which would be LL in this case). Conversely, even the lowest order
FONLL matching, FONLL-A is NLO+NNLL\cite{Forte:2015hba,Forte:2018ovl}.
Nevertheless, the multijet-merging method is systematically
improvable, by either including higher fixed order corrections, or
corrections to the parton shower.
Some relevant distributions are shown in Fig.~\ref{fig:mjm}, for the
exampled described above.

An somewhat older and alternative approach, was first introduced in
\cite{Krauss:2017wmx}, and consists in extending the 5FS scheme to
include mass effects even for initial state particles. This has the
advantage to include the resummation of high energy logs, which are
included in the $b$-PDF, as well as mass power corrections, which then
enter in the matrix elements and phase-space. A the time of
\cite{Krauss:2017wmx}, however, this method suffered by two important
limitations. First, it
is known that calculations with initial state massive quarks suffer
from higher-order uncancelled soft-divergence, which may render the
NLO result useless. Second, it was not clear whether or not one was
allowed to consistently use standard PDFs with massive initial states. 
As for the former issue, research to study to what extent this is an
actual issue are ongoing.

The latter issue has instead recently been
resolved~\cite{Forte:2019hjc}. Indeed it has been shown that a 5F
massive scheme\footnote{Note that this is true for any heavy flavour
  scheme.} can be consistently construct by simply constructing a
parametrised $b$-PDF, and performing the FONLL matching between the
4FS and the 5FS: the matched FONLL result is indeed identical to what
is defined in Ref.~\cite{Krauss:2017wmx}.
\begin{figure}[t]
  \centering{
    \includegraphics[width=0.45\linewidth]{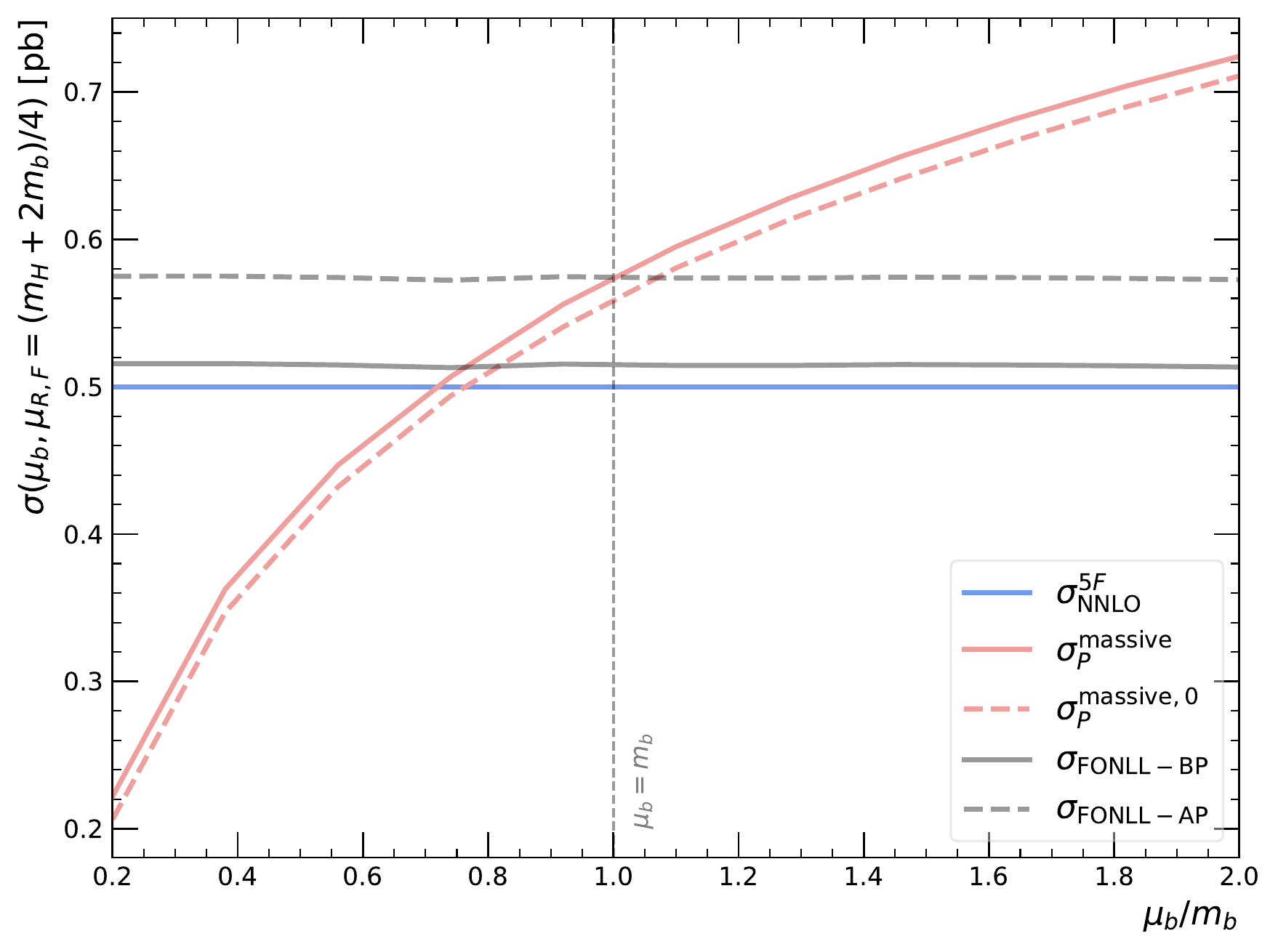}
    \includegraphics[width=0.45\linewidth]{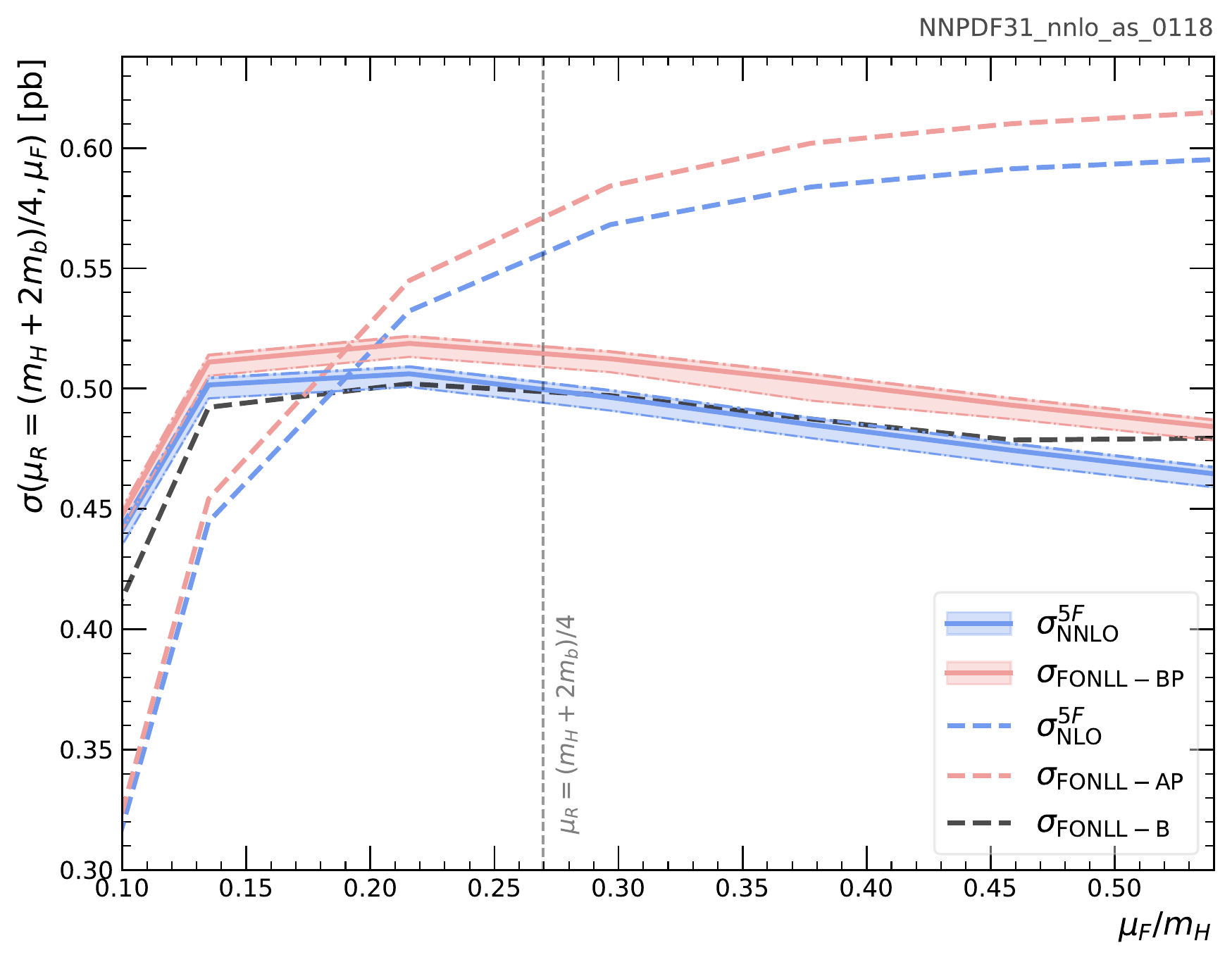}
  }
  \caption{\label{fig:fonll-bp} Results in the FONLL scheme, and
    comparison to the standard 5FS.}
\end{figure}
The caveat is that one has to assume that, in some sense, some
intrinsic component of the heavy flavour exists. However, this can be
simply seen by lowering the heavy quark threshold. This already
happens in standard PDF sets, where the $b$-PDF is given by matching
conditions: starting at NNLO, constant terms in the matching
conditions make the heavy quark PDF non-zero even below threshold.

Just as an example, we report in Fig.~\ref{fig:fonll-bp} results
obtained for $b\bar{b}\to H$ in this scheme. On the left panel we show
the independence of the matched cross section with respect to the
heavy quark matching scale, while on the right hand side a comparison
between the 5FS and the 5F massive (or FONLL) is shown as a function
of the factorisation scale.

In conclusion, in this talk we have presented some recent theoretical
developments needed to achieve a more accurate description of
processes involving heavy flavours. This is, in particular, related to how
one decides to treat initial state heavy quarks.
Improvements at the inclusive and at the differential level are
on-going. While the multi-jet merging approach presents various
interesting aspects, it still needs some detail studies and comparisons,
at least in the view of the authors.
On the other hand recent theoretical progresses in the matching of
inclusive calculations, have shown to what extent the use of a massive
heavy flavour scheme, or in other words a scheme in which heavy
flavours are taken as massive even when in the initial state, is
reliable. We hope that in the future fitted $b$-PDFs will be provided
together with standard PDFs, as this will make the use of such schemes
completely consistent.

\end{document}

%% file: tikz.tex
\usepackage{
  pgf,
  tikz}
\usepackage{newtxsf}
\usetikzlibrary{
  arrows,
  trees,
  scopes,
  decorations.text,
  decorations.pathreplacing,
  decorations.pathmorphing,
  decorations.markings,
  positioning,
  calc,
  patterns,
  intersections,
  shapes,
  fadings,
  matrix
}

\tikzset{
  photon/.style=
  {
    decorate,
    decoration={snake},
    draw=black
  },
  particle/.style=
  {
    very thick,
    draw=black,
    postaction={decorate},
    decoration=
    {
      markings,
      mark=at position .5 with {\arrow[draw=black]{>}}
    }
  },
  lparticle/.style=
  {
    draw=black,
    postaction={decorate},
    decoration=
    {
      markings,
      mark=at position .5 with {\arrow[draw=black]{>}}
    }
  },
  antiparticle/.style=
  {
    very thick,
    draw=black,
    postaction={decorate},
    decoration=
    {
      markings,
      mark=at position .5 with {\arrow[draw=black]{<}}
    }
  },
  gluon/.style=
  {
    decorate,
    draw=black,
    decoration=
    {coil,
      amplitude=4pt,
      segment length=5pt
    }
  },
  psgluon/.style=
  {
    decorate,
    draw=black,
    decoration=
    {coil,
      amplitude=2pt,
      segment length=5pt
    }
  },
  scalar/.style=
  {
    densely dashed,
    draw=black
  },
}
\tikzset{
  hatch distance/.store in=\hatchdistance,
  hatch distance=10pt,
  hatch thickness/.store in=\hatchthickness,
  hatch thickness=0.3pt,
}
\makeatletter
\pgfdeclarepatternformonly[\hatchdistance,\hatchthickness]{north west hatch}
{\pgfqpoint{0pt}{0pt}}
{\pgfqpoint{\hatchdistance}{\hatchdistance}}
{\pgfpoint{\hatchdistance-1pt}{\hatchdistance-1pt}}%
{
  \pgfsetcolor{\tikz@pattern@color}
  \pgfsetlinewidth{\hatchthickness}
  \pgfpathmoveto{\pgfqpoint{\hatchdistance}{0pt}}
  \pgfpathlineto{\pgfqpoint{0pt}{\hatchdistance}}
  \pgfusepath{stroke}
}
\makeatother

%% file: skeleton.bbl
\begin{thebibliography}{99}
\bibitem{Forte:2015hba}
  S.~Forte, D.~Napoletano and M.~Ubiali,
  Phys.\ Lett.\ B {\bf 751} (2015) 331
  doi:10.1016/j.physletb.2015.10.051
  [arXiv:1508.01529 [hep-ph]].
\bibitem{Forte:2016sja}
  S.~Forte, D.~Napoletano and M.~Ubiali,
  Phys.\ Lett.\ B {\bf 763} (2016) 190
  doi:10.1016/j.physletb.2016.10.040
  [arXiv:1607.00389 [hep-ph]].
\bibitem{Krauss:2016orf}
  F.~Krauss, D.~Napoletano and S.~Schumann,
  Phys.\ Rev.\ D {\bf 95} (2017) no.3,  036012
  doi:10.1103/PhysRevD.95.036012
  [arXiv:1612.04640 [hep-ph]].
\bibitem{Forte:2018ovl}
  S.~Forte, D.~Napoletano and M.~Ubiali,
  Eur.\ Phys.\ J.\ C {\bf 78} (2018) no.11,  932
  doi:10.1140/epjc/s10052-018-6414-8
  [arXiv:1803.10248 [hep-ph]].
\bibitem{Bagnaschi:2018dnh}
  E.~Bagnaschi, F.~Maltoni, A.~Vicini and M.~Zaro,
  JHEP {\bf 1807} (2018) 101
  doi:10.1007/JHEP07(2018)101
  [arXiv:1803.04336 [hep-ph]].
\bibitem{Hoche:2019ncc}
  S.~H\"oche, J.~Krause and F.~Siegert,
  Phys.\ Rev.\ D {\bf 100} (2019) no.1,  014011
  doi:10.1103/PhysRevD.100.014011
  [arXiv:1904.09382 [hep-ph]].
\bibitem{Krauss:2017wmx}
  F.~Krauss and D.~Napoletano,
  Phys.\ Rev.\ D {\bf 98} (2018) no.9,  096002
  doi:10.1103/PhysRevD.98.096002
  [arXiv:1712.06832 [hep-ph]].
\bibitem{Forte:2019hjc}
  S.~Forte, T.~Giani and D.~Napoletano,
  Eur.\ Phys.\ J.\ C {\bf 79} (2019) no.7,  609
  doi:10.1140/epjc/s10052-019-7119-3
  [arXiv:1905.02207 [hep-ph]].
\end{thebibliography}
